# A proposal for a revised meta-architecture of intelligent tutoring systems to foster explainability and transparency for educators

Florian Gnadlinger[1,2] [a] and Simone Kriglstein[2] [b]
[1] *Faculty of Computer Science, Communication, and Economics, University of Applied Sciences Berlin, Germany*
[2] *Faculty of Informatics, Masaryk University, Czech Republic*
*florian.gnadlinger@htw-berlin.de, kriglstein@mail.muni.cz*

Abstract: This contribution draws attention to implications connected with meta-architectural design decisions for intelligent tutoring systems in the context of formative assessments. As a first result of addressing this issue, this contribution presents a meta-architectural system design that includes the role of educators.

## 1 INTRODUCTION

According to an ongoing systematic literature review and similar reviews that have already been conducted [1–3], we see evidence that intelligent tutoring systems are developed from a conceptual perspective in a black-box manner for the actual participants of educational assessment scenarios.

This emphasizes the necessity to examine system designs, employed algorithms, and the conceptual implementation of pedagogical-psychological models in intelligent tutoring systems to enhance explainability and transparency for all stakeholders. A first step to do so is to reflect on intelligent tutoring systems beyond their current borders and to question:

*RQ1: How can we incorporate the role of educators into meta-architectural designs of intelligent tutoring systems to foster their explainability and transparency?*

## 2 BACKGROUND

Competency-based learning (or competency-based education and related synonyms as discussed in [4]) refers to a pedagogical approach that supports the development of practical skills and behaviors that are necessary for success in real-world situations [5, 6]. A revised definition of competency-based learning distinguishes this concept into seven major aspects [6, 7]. Two of these aspects address the assessment of competencies: "…(2) assessment is meaningful, timely, relevant, and actionable evidence; … (4) students' assessment is based on evidence of mastery; [6, p.1904]"(retrieved from Levine & Patrick [7]).

Formative assessments are a major form of assessing competencies [8–11]. Usually, these kinds of assessments are typically incremental [9, 10, 12] and inform not only learners about their current state but also educators about the effectiveness of their pedagogical and didactical methods [11]. To design, develop, and implement formative assessments, various methodologies have been developed to derive these information for educators, for example, the ADDIE Model [13], Four – Component Instructional Design (4C/ID) [14] or the Evidence Centered-Design Framework (ECD) [15].

The ECD is based on the idea that when learners fulfill tasks, they create some kind of result (*work product*) that incorporates, to some degree, the learner's competence level [16]. Hence, extracting *evidence* from the work products within a defined evidence identification process leads to *claims* about certain competencies [15, 17, 18]. Collecting and storing these beliefs result in an individual description of each learner, a *learner model*. Thus, before constructing the actual assessment, it is necessary to reflect on all those stated parts and other aspects as well [15].

Educators usually act in a similar (but far less formal) way to develop and apply formative assessments in their educational settings, which costs them a lot of time. A recent McKinsey study ([19]) with a focus on K-12 teachers in the US, UK, CA, and SG reveals that they spend 34% of their time on preparation, evaluation, and feedback tasks [19, 20].

Hence, researchers and developers have been eager to tailor computer-based learning experiences to the

---

[a]  https://orcid.org/0000-0003-4569-9671 (main author)
[b]  https://orcid.org/0000-0001-7817-5589 (review & organisation)

learner´s needs by using evidence about competencies to adapt the learning process autonomously [21–23]. To achieve this, such systems incorporate *learner models*, *task* (or more general *domain*) *models*, and automated evidence identification processes and use those to generate adaption recommendations. Examples of conceptional formulations for such systems could be recommending the next best-fitting learning object (e.g., a task with a specific difficulty), finding a sequence of novel courses, or supporting the search for learning peers [24, 25].

Nowadays, advanced systems are able to mimic the behaviors of human tutors by using artificial intelligence, which results in their classification as intelligent tutoring systems [26, 27]. To do so, these systems not only incorporate *learner models* and *domain models* but also so-called *tutor models*. The tutor model is the decision entity that schedules pedagogical or didactical interventions according to the state of the learner model as a response to the learner's interaction [28]. As a result of this perspective, intelligent tutoring systems are designed and developed according to the meta-architectural approach illustrated by the gray elements in Figure 1 (see [26, 28–30]).

## 3 METHODE & RESULTS

The presumption to the given research question (RQ1) is that educators are only able to use the full potential of intelligent tutoring systems if they (1) have access to the information obtained from the learners, (2) are able to understand and interpret this information, (3) and can transform this interpretation into valuable pedagogical and didactical actions. This aligns with the learning analytics process model [31] (see Figure 2), which is applicable to learner or teaching dashboards and authoring interfaces.

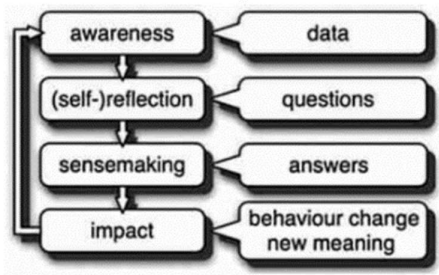

Figure 2: learning analytics process model [31].

Hence, regarding the results from an ongoing systematic literature review, we concluded a revised meta-architecture of intelligent tutoring systems that incorporate the role of educators (Figure 1 black elements). This draws attention to the design of teaching dashboards, allowing *views into* and *interactions with* the different models of such systems (compare with Figure 1). Besides static dashboards, a functional entity is needed to support the educators´ reflection process about the effectiveness of their teaching methods. We call this entity educator model.

## 4 DISCUSSION

With the given summary, the illustrated current systematic overview (Figure 1 gray elements), and a visualized extension proposal (Figure 1 black elements), we would like to point out a major implication for teachers in higher education when introducing intelligent tutoring systems into their educational setting.

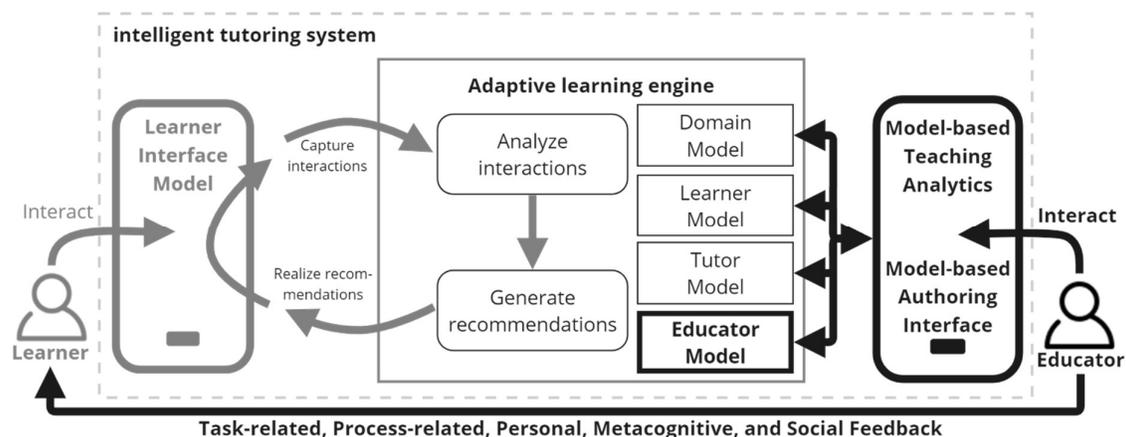

Figure 1: Adaptive learning loop within digital learning environments. All gray elements represent the state-of-the-art system architecture of intelligent tutoring systems. All black elements represents the proposed extension.

If teachers in higher education are using or will start using intelligent tutoring systems, they should reflect on three main questions.

(1) Do I have access to all information incorporated into the different models of intelligent tutoring systems?

(2) I am able to understand and interpret this information?

(3) I am able to transform this interpretation into pedagogical or didactical actions?

## 5 CONCLUSION

With this contribution, we hope to give higher education teachers some leverage to participate in the discussion of the design of intelligent tutoring systems.